\documentclass{aa}
\usepackage{natbib,amssymb}
\usepackage[pdftex]{graphicx,color}
\usepackage{rotating,footmisc}
\bibpunct{(}{)}{;}{a}{}{,}
\newcommand{\solm}{M$_{\odot}$}

\title{IRS$\,$13N: a new comoving group of sources at the Galactic Center}
\subtitle{}
 \author{K. Mu\v{z}i\'{c}$^{1,2}$, R. Sch\"odel$^3$, A. Eckart$^{1,2}$, L. Meyer$^4$, A. Zensus$^{2,1}$} 
\institute{1) I. Physikalisches Institut, Universit\"at zu K\"oln,
           Z\"ulpicher Str. 77,
           50937 K\"oln, Germany\\
           2) Max-Planck-Institut f\"ur Radioastronomie, 
           Auf dem H\"ugel 69, 
	   53121 Bonn, Germany\\
           3) Instituto de Astrof\'isica de Andaluc\'ia,
              Camino Bajo de Huétor 50,
              18008 Granada, Spain \\
           4) University of California, Division of Astronomy and Astrophysics,
	      Los Angeles, CA 90095-4705\\
           \email{muzic@ph1.uni-koeln.de} }

\date{Received  / Accepted }
\begin{document}
\abstract
{The Galactic Center IRS$\,$13E cluster is located $\sim$3.2'' from SgrA*. It is an extremely
dense stellar association containing several Wolf-Rayet and O-type stars, at least four of which show
a common velocity. Only half an arcsecond north from IRS$\,$13E 
there is a complex of extremely red sources, so-called IRS$\,$13N. Their nature is 
still unclear. Based on the analysis 
of their colors, there are two main possibilities: 
(1) dust embedded sources older than few
Myr, or (2) extremely young objects with ages less than 1Myr.}
{We present the first proper motion measurements of IRS$\,$13N members, and additionally give 
proper motions of four of IRS$\,$13E stars resolved in the L'-band.}
{The L'-band (3.8$\mu$m) observations have been carried out using the NACO adaptive
 optics system at the ESO VLT.
 Proper motions
have been obtained by linear fitting of the stellar positions 
extracted by StarFinder as a function of time, weighted by positional uncertainties.}   
{We show that six of seven resolved northern sources show a common proper motion, thus revealing 
a new comoving group of stars in the central half parsec of the Milky Way. The common proper motions
of IRS$\,$13E and IRS$\,$13N clusters are significantly ($>$5$\,$$\sigma$) different. 
We also performed 
a fitting of the positional data for those stars onto
Keplerian orbits, assuming SgrA* as the center of the orbit. Our results favor
the very young stars hypothesis.}
{}

\authorrunning{K. Mu\v{z}i\'{c} et. al.} 
\titlerunning{IRS13N comoving group}

\maketitle


\section{Introduction}
\label{intro}

The central half parsec of the Milky Way is host to
 a surprisingly high number of massive young stars 
(see e.$\,$g. \citealt{paum06,ghez05}), organized
in at least one disk-like structure
of clockwise rotating stars (CWS; \citealt{genzel03,l&b03,paum06}).
\citet{paum06} also propose the existence of a second, less populated
disk of counter-clockwise rotating stars (CCWS).
The mechanism responsible for the presence of young stars in the strong
tidal field of the super-massive black hole (SMBH) at the position of 
SgrA* is not clear. Two most prominent scenarios include star formation
``in-situ'' (in an accretion disk; \citealt{l&b03, nayakshin06a}), 
and the in-spiral of a massive 
stellar cluster formed at a safe distance of 5-30 pc 
from the Galactic Center (GC; \citealt{gerhard01,mcm&p-z03,kim04,p-z06}). 
At this point, the former scenario seems to be favored
by number of authors \citep{nayakshin&sunyaev05,nayakshin06a,paum06}.
Also, recent results by \citet{stolte07} definitively rule out
the possibility that the Arches cluster could migrate inwards and fuel
the young stellar population in the GC.

A special challenge and a good observational test for both the above mentioned
scenarios is provided by the existence of the IRS 13 group of sources. 
IRS 13E (located $\sim$$\,$3'' west and $\sim$$\,$1.5'' south of SgrA*)
is the densest stellar association after the stellar cusp
in the immediate vicinity of SgrA* and contains several 
massive Wolf-Rayet (WR) and O-type stars \citep{maillard04, paum06, moultaka05}. 
It is generally considered to be associated with the mini-spiral material
\citep{moultaka05, paum04} and is probably bound, since four out of seven 
identified stars show highly correlated velocities \citep{maillard04, schoedel05}.   
For both of the above mentioned scenarios of star formation at the GC there are
several issues when dealing with IRS\,13E. 
In principle, such a cluster could have been formed in an accretion disk 
\citep{milos&loeb04, nayakshin05}. However, 
in numerical simulations of star forming disks, fragmentation of a disk 
cannot produce such a dominant feature \citep{nayakshin07}. 
In light of the cluster in-fall scenario,
an intermediate-mass black hole (IMBH) was proposed to reside in
the center of the cluster \citep{maillard04}. 
The existence of an IMBH makes the process of
cluster in-spiral much more efficient in terms of increasing
dynamical friction and thus allowing the most massive stars that reside
in the center of a very massive ($>$10$^{6}$\solm) stellar cluster to reach the central
parsec of the Galaxy within their lifetimes (\citealt{hansen&milos03,berukoff06, p-z06}; 
but see \citealt{kim04} for a characterization of the problems with this hypothesis).
However, the presence
of the IMBH in IRS 13E is disputed. \citet{schoedel05} analyze the
velocity dispersion of cluster stars and derive that the mass
of such an object should be at least 7000$\,$\solm,  
 making its existence implausible. 
Both the X-ray \citep{bag03} and
radio \citep{zhao&goss98} source at the position of IRS 13E can be
explained by colliding winds of high-mass losing stars \citep{coker02,zhao&goss98}, 
without the need for any unusual object.
\citet{paum06} suggest that, in the case that IRS 13E is associated with
the Bar of the mini-spiral, the mass of the stellar content would be high enough
to keep the cluster from disruption. 

A second comoving group of stars, called IRS$\,$16SW and located 1.9'' 
in projection from SgrA*, was reported by \citet{Lu05}. 
Four of the five members of the IRS$\,$16SW comoving group 
are spectroscopically identified as young massive stars, indicating ages comparable
to those of the IRS$\,$13E cluster. However, unlike IRS$\,$13E,
IRS$\,$16SW shows only a slight stellar
number density enhancement.
 
Approximately 0.5'' north of IRS 13E, a small cluster of unusually
red compact sources has been reported (IRS\,13N; \citealt{eckart04a}). Eight sources
have been resolved and labeled $\alpha$ through $\eta$ as shown
 in Figs. 1 and 2 of \citet{eckart04a}.
A strong infrared excess is due to the emission of warm (T $\sim$$\,1000$$\,$K) dust 
\citep{moultaka05}. 
The authors
propose two possible explanations for the nature of IRS 13N: 
(1) objects older than few Myr and similar to bow-shock sources 
reported by \citet{tanner05} or (2) extremely young stars (0.1 - 1 Myr old).
The latter scenario implies more recent star formation than what has been
assumed so far for the GC environment. 

Here we present first proper motion measurements of the IRS$\,$13N sources and 
additionally present proper motions of four of IRS$\,$13E sources, including
one for which no proper motion has previously been published. All proper motions
 were derived using L'-band NACO/VLT images
\footnote{Based on observations collected at the European Southern Observatory, Chile}.

  \begin{figure*}
\centering
 \resizebox{17cm}{!}{\includegraphics{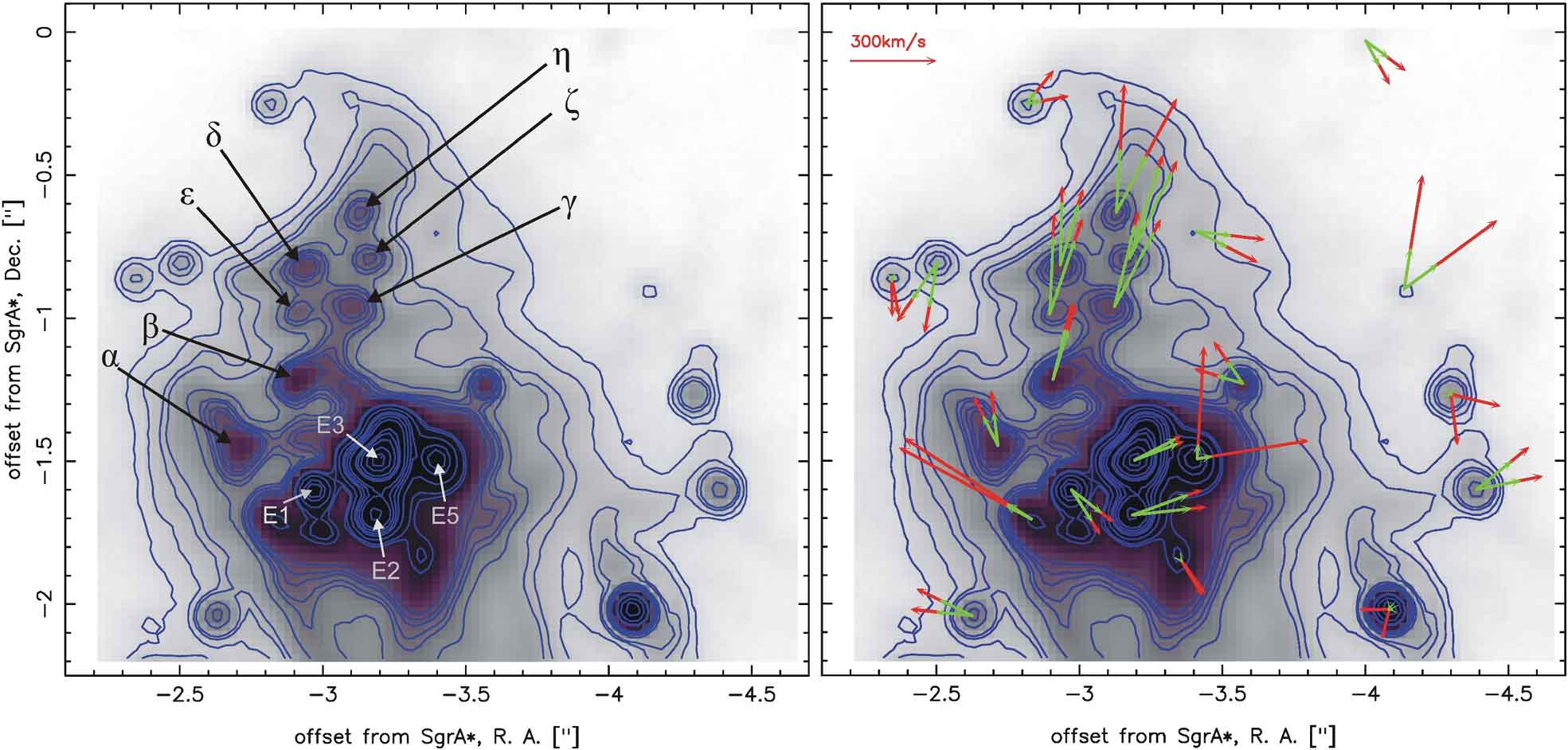}}
 \caption{Identification and proper motions of stars in IRS$\,$13 complex. Four
arrows are shown for each source, indicating the $\pm$3$\,$$\sigma$ uncertainty 
of the value and direction of its proper motion.}
\label{promot}
\end{figure*}

\begin{table}
\caption{Proper motions of IRS$\,$13N sources observed in L'-band.} 
\label{13Ntable}
\begin{center}
\begin{minipage}{1.0\textwidth}
\begin{tabular}{lccrrrr}
\hline\hline
name & 
$\Delta$$\alpha$ \footnote{relative to SgrA*, in arcseconds \label{fn3}} & 
$\Delta$$\delta$ \footref{fn3} &
$v_{R. A.}$ \footnote{all velocities are in km$\,$s$^{-1}$; the uncertainties \\
represent the 
1$\sigma$ uncertainty of the linear fit \label{fn6}} & 
&
$v_{Dec.}$ \footref{fn6}& 
\\
\hline

\end{tabular}\\
\begin{tabular}{c c cr@{ $\pm$}rr@{ $\pm$}r}

$\alpha$  &  -2.70 & -1.48 &   44  &  8 & 147 & 13\\
$\beta$   &  -2.89 & -1.25 &  - 6  & 20 & 228 & 15\\
$\gamma$  &  -3.10 & -0.99 & -114  & 14 & 298 &  7\\
$\delta$  &  -2.92 & -0.85 &  -33  & 13 & 248 &  9\\
$\epsilon$&  -2.88 & -1.02 &  -58  & 15 & 306 & 13\\
$\zeta$   &  -3.16 & -0.82 & -134  &  9 & 333 &  6\\
$\eta$    &  -3.11 & -0.66 &  -91  & 22 & 323 & 39\\
13N\footnote{average proper motion of six comoving sources $\beta$-$\eta$}  &  -3.01 & -0.93 &  -73  &  7 & 289 &  8\\

\end{tabular}
\end{minipage}
\end{center}
\end{table}
 
\begin{table}
\caption{Proper motions of IRS$\,$13E sources observed in L'-band.} 
\label{13Etable}
\begin{center}
\begin{minipage}{1.0\textwidth}
\begin{tabular}{lccrrrr}
\hline\hline
name & 
$\Delta$$\alpha$ \footnote{relative to SgrA*, in arcseconds \label{fn2}}& 
$\Delta$$\delta$ \footref{fn2} &
$v_{R. A.}$ \footnote{all velocities are in km$\,$s$^{-1}$; the uncertainties\\
 represent the 
1$\sigma$ uncertainty of the linear fit \label{fn4}} & 
& 
$v_{Dec.}$ \footref{fn4}& \\
\hline
\end{tabular}\\
\begin{tabular}{cccr@{ $\pm$}rr@{ $\pm$}r}

E1  &  -2.95 & -1.64 & -107 & 11 & -118 & 6\\
E2  &  -3.16 & -1.73 & -228 &  8 &   53 & 9\\
E3  &  -3.17 & -1.53 & -158 &  4 &   71 & 3\\
E5  &  -3.39 & -1.53 & -175 & 45 &  140 & 72\\
13E \footnote{average proper motion of four IRS$\,$13E sources}& -3.17 & -1.61 & -167 & 12 & 37 & 18\\
13E \footnote{average proper motion of IRS$\,$13E without E1}&-3.24 & -1.60 &-187 & 15 & 88 & 24\\

\end{tabular}
\end{minipage}
\end{center}
\end{table}
 
\section{Observations and Data Reduction}
\label{reduction}                 

The L' (3.8 $\mu$m)
images were taken with the NAOS/CONICA adaptive optics assisted imager/spectrometer 
\citep{lenzen98,rousset98,brandner02}
at the UT4 (Yepun) at the ESO VLT.
The data set includes images from 6 epochs 
(2002.66, 2003.36, 2004.32, 2005.36, 2006.41 and 2007.39) 
with a resolution of $\sim$100$\,$mas and a pixel scale of 27$\,$mas/pixel.
Data reduction (bad pixel correction, sky subtraction, flat field correction) 
and formation of final mosaics was performed using the DPUSER software for 
astronomical image analysis (T. Ott; see also \citealt{eckart90}).  
All the images were deconvolved using the linear Wiener filter. 
The absolute positions of sources in our AO images were derived by comparison to the
VLA positions of IRS~10EE, 28, 9, 12N, 17, 7 and 15NE as given by \citet{reid03}.

Stellar positions were extracted with StarFinder \citep{diolaiti00} and transformed
into the common coordinate system with the aid of 19 reference stars. The positions 
of the reference stars 
were corrected for the stellar proper motions as derived from the K$_S$-band images 
(see \citealt{muzic07}).  
In \citet{muzic07} we have shown that the lower resolution L'-band data
can be reliably used to obtain proper motions. 

Proper motions were derived by linear fitting of positions as a function
of time, weighted by the positional uncertainties. Both the error of 
the transformation to the reference frame and the error of the stellar
position fitting contribute to the uncertainties.

We assume the distance to the GC to be 7.6$\,$kpc \citep{eisenhauer05}.

\section{Results}
\label{results}                 

\subsection{Proper motions}
\label{pm}

In tables \ref{13Ntable} and \ref{13Etable} we list proper motions of 
IRS$\,$13N and IRS$\,$13E sources identified in L'-band, respectively.
In Fig.$\,$\ref{promot} we show proper motions of all stars
superposed on the L'-band image. 
Stars $\beta$ through $\eta$ show similar proper motions,
revealing a new comoving group of sources at the GC. 
As shown in \citet{eckart04a}, only $\eta$ can also be detected in
the K-band.  
\citet{paum06} give the full velocity information for this star: 
$v$$_{R.A.}$$\,$=$\,$(-52$\pm$28)kms$^{-1}$, 
$v$$_{Dec.}$$\,$=$\,$(257$\pm$28)kms$^{-1}$ and 
$v$$_{r}$$\,$=$\,$(40$\pm$40)kms$^{-1}$. The authors identify
it as a member of a possibly existing second stellar disk (CCWS).

Concerning IRS$\,$13E, the obtained proper motion for E1 shows 
the biggest discrepancy with respect
to previously published data \citep{paum06,schoedel05}. 
It also deviates significantly from the proper motions of the other three stars.
It was already noted that E1 shows the largest deviation from the systemic proper motion
of the cluster \citep{schoedel05}. \citet{paum06} argue that this star is possibly
not bound to the cluster. This is supported by our result.

It is important to note that, in addition to a very dense stellar population
in the region, a large amount of
diffuse dust emission detected at 3.8$\,$$\mu$m makes the astrometry more challenging
to perform than at shorter wavelengths.
For this reason, StarFinder was not able to resolve all the 
previously reported K-band sources in the region. E3 is known to be at least
a double source, while here we present it as a single one. 
E4 is located close to E3 but is much fainter
in the L'-band. Therefore it could not be clearly 
identified by StarFinder. 
However, we were able to resolve the faint source E5
in all epochs and present the first proper motion measurement 
of this star.
E5 was proposed to be a dusty WR star 
\citep{maillard04}, and it seems to exhibit a proper 
motion consistent with those of E2 and E3. This implies that it may be a cluster member.

From Tables \ref{13Ntable} 
and \ref{13Etable} one can see that the
average proper motions of the two clusters are significantly ($>$$\,$5$\,$$\sigma$)
different. 
Thus, it seems that two clusters do not belong to the same system. 

\subsection{Keplerian orbit fitting}
\label{kepler}

  \begin{figure*}
\centering
 \resizebox{15cm}{!}{\includegraphics{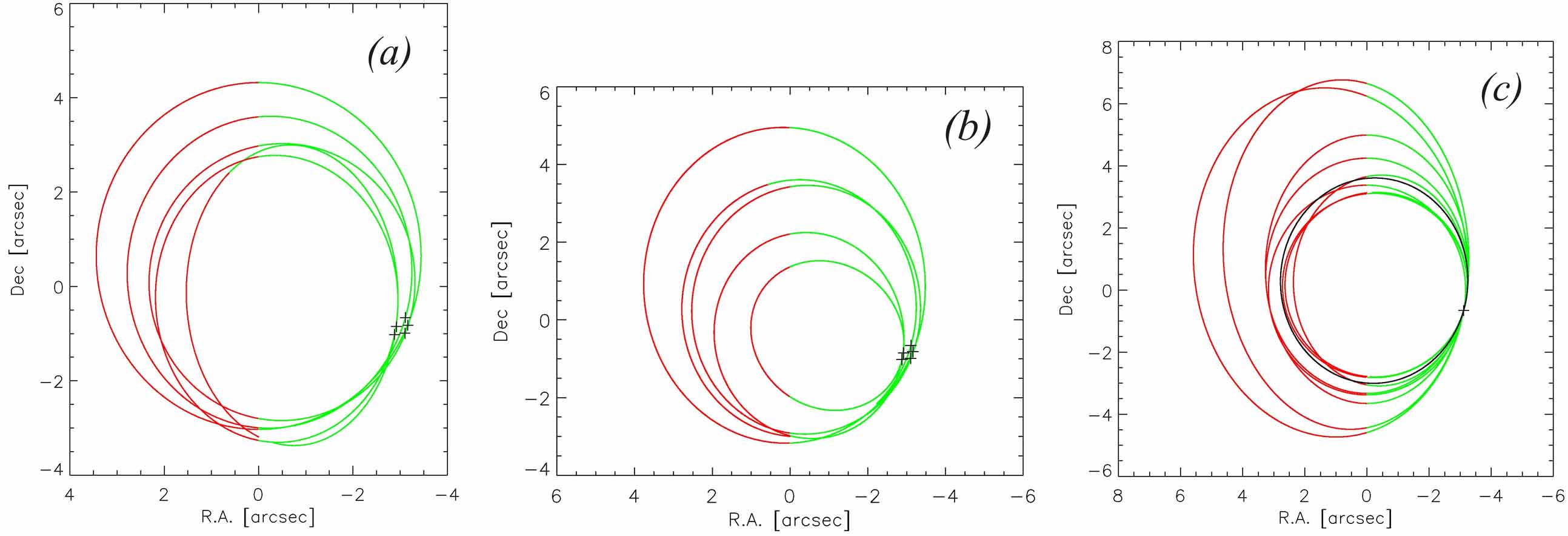}}
 \caption{ $(a)$ the best fit orbital solutions for five IRS$\,$13N stars ($\gamma$ -- $\eta$);
           $(b)$ orbital solutions for a single plane ($i$=24$^o$, $\Omega$=180$^o$), for stars $\gamma$ -- $\eta$ ; 
           $(c)$ chosen 1$\sigma$ orbits for $\eta$, with the best fit orbit colored black;
NOTES: part of the orbit in front of the plane of the sky is colored red; crosses mark present positions of the stars; axes show the offset from SgrA*.}
\label{orbits}
\end{figure*}

In order to analyze our measurements we attempted to fit
the positional data of IRS$\,$13N stars showing similar proper motions
 (stars $\beta$ to $\eta$)
 to Keplerian orbits, assuming 
that the gravitational potential is dominated by the
 SMBH at the position
of SgrA*. For details of the fitting procedure and definition of
 parameters see e.g. \citet{eisenhauer03}. We assumed the distance and the
mass of SgrA* to be 7.6$\,$kpc and 3.6$\times$10$^6$$\,$\solm, respectively.
The velocity of SgrA* was assumed to be zero \citep{reid04}.
Under the assumption that the stars are really on bound orbits around SgrA*, 
it is clear that our data cover only a small part of such an orbit. We also
lack the radial velocity information for all of the stars. As a consequence,
the $\chi$$^2$ minimization gives a very wide span of orbital 
parameters for which $\chi$$^2$$<$$\chi$$^2$$_{min}$+1, even when fixing most of the free 
parameters (the mass of and the 
distance to SgrA*, as well as the position and the velocity of 
the center of the orbit). 
Therefore we repeated the orbital fitting for each star introducing
the radial velocity data as given in \citet{paum06} for $\eta$. 
We varied the inclination angle $i$ between 0$^o$ and 90$^o$ in steps of 1$^o$, 
$\Omega$ (position angle of the ascending node) between 0$^o$ and 180$^o$ in steps of 20$^o$
and $\omega$ (longitude of periastron) between 0$^o$ and 360$^o$ with a step size of 20$^o$.
The results are given in Table \ref{Kepler}. We list the best fit parameters $i$ and $\Omega$, 
and give the 1$\sigma$ range for the inclination.
We also list the range in eccentricities ($e$) and semi-major axes ($a$) 
resulting from the fits for which there is some
 combination of angles ($i$, $\Omega$, $\omega$) that gives a fit 
better than 1$\sigma$
(\footnote{a fit with $\chi$$^2$$<$$\chi$$^2$$_{min}$+1, 
for a corresponding star. For a convenience,
we refer to all orbits that satisfy this condition as 1$\sigma$ orbits.}\label{foot2}).
For all the stars except $\zeta$, the best fit reduced
$\chi$$^2$ value ranges between 0.6 and 1.5. For $\zeta$ the best fit
is obtained with $\chi$$^2$=3.9, probably implying that either positional uncertainties
are underestimated, or that assuming the radial velocity of $\eta$ is a poor assumption. 
In the last three lines of Table \ref{Kepler} we give the best-fit
parameters for the Keplerian orbits fitted onto the average positions of
the stars within the IRS13N and IRS13E comoving groups.
For IRS$\,$13E we perform the fitting procedure twice: first including the
star E1 and afterwards considering it a non-member. 

\citet{paum06} conclude that IRS$\,$13E is on a fairly eccentric orbit, with $e$$\gtrsim$0.5,
which we confirm.
For all IRS$\,$13N stars, with the exception of $\beta$, the range in
 eccentricities of 1$\sigma$ orbits is restricted to fairly low values ($<$$\,$0.4),
in agreement with the \citet{paum06} value listed for $\eta$. Note that the corresponding semi-major axes
also result fall within a restricted range of values. In contrast, both parameters ($e$ and $a$) 
are poorly constrained for the star $\beta$. The size of the best fit orbit
of $\beta$ is an order of magnitude higher than those of the other five stars. 
Since the orbit of $\beta$ is radically different from orbits of all the other stars, 
we note that it probably should not be regarded as part of the co-moving group. Therefore we do not show
its orbit in Fig.$\,$\ref{orbits}.
Since $\eta$ is the only star with known radial velocity, we took its orbital parameters
and attempted to fit other stars onto the same orbit. 
We note that it is not possible to fit all other stars onto exactly the same orbit 
(same $i$, $\Omega$ and $\omega$).
Therefore we fix $i$ and $\Omega$ and let $\omega$ vary. This confines the stars to the same plane, 
within which they can still have different orbits.
For three stars
($\gamma$, $\epsilon$, $\zeta$) there exist 1$\sigma$ orbits with given $i$ and $\Omega$, 
the star $\delta$ can be reasonably
(but still not within 1$\sigma$) fitted, and $\beta$ fails completely to be fitted onto this plane
(Fig.$\,$\ref{orbits}$b$).
To get a feeling about the uncertainties of our analysis,
we plot several 1$\sigma$ orbits for the same star ($\eta$) in Fig.$\,$\ref{orbits}$c$. 

The implications of these results are further discussed in Section \ref{bound}.

\begin{table*}
\begin{center}
\caption{Results of the Keplerian orbit fitting} 
\label{Kepler}
\begin{minipage}{1.0\textwidth}
\begin{tabular}{cccccc}
\hline\hline
name & 
$i$($^o$)\footnote{best fit parameters ($\chi$$^2$=$\chi$$^2$$_{min}$) \label{fn5}} & 
$\Omega$($^o$)\footref{fn5} &
$i$($^o$)\footnote{all inclination values for which 
$\chi$$^2$$<$$\chi$$^2$$_{min}$+1 \label{fn7}} &
$e$\footnote{calculated for ($i$, $\Omega$, $\omega$) combinations for which $\chi$$^2$$<$$\chi$$^2$$_{min}$+1 \label{fn8}}&
$a$($''$)\footref{fn8}\\
\hline
$\beta$   &  79 &  180 &  53 - 80 & 0 - 0.9 & 4.0 - 180.0\\
$\gamma$  &  19 &  180 &  4 - 67  & 0.17 - 0.21 & 2.6 - 4.0\\
$\delta$  &  48 &  180 &  39 - 71 & 0.15 - 0.4  & 2.7 - 15.0\\
$\epsilon$&  33 &  180 &  17 - 56 & 0.1 - 0.3   & 2.5 - 7.2\\
$\zeta$   &  18 &  180 &  0 - 30  & 0.1 - 0.3   & 3.2 - 4.7\\
$\eta$    &  24 &  180 &  1 - 45  & 0.1 - 0.3   & 2.8 - 5.7\\
IRS$\,$13N \footnote{average orbit \label{fn9}}&  28 &  180  &  14 - 49 & 0.1 - 0.2  &  2.8 - 5.7\\
IRS$\,$13E \footref{fn9}$^,$\footnote{including E1}&  83 &  100  &  80 - 85 & 0.5 - 0.9  &  12.5 - 145.0\\
IRS$\,$13E \footref{fn9}$^,$\footnote{not including E1}&  70 &  120 &  14 - 42; 58 - 80 &  0.5 - 0.8  &  2.3 - 66.0 \\
\end{tabular}
\end{minipage}
\end{center}
\end{table*}

\section{Discussion} 
\label{disc}

\subsection{3-dimensional position}
\label{3d}

\citet{paum04} report the morphology and radial velocities of the mini-spiral
gas based on He~I and Br~$\gamma$
observations. At the position of the IRS$\,$13 sources, the Northern Arm and the Bar
of the mini-spiral are overlapping along the line of sight, with the Bar
being placed further away from the observer. 
The authors suggest that IRS$\,$13E must be either embedded in the Bar,
or very close to it. Also, the radial velocities of the IRS$\,$13E stars are similar to
the radial velocities of the Bar material at this position \citep{paum04}. 
Spectroscopic observations of the entire IRS$\,$13 region \citep{moultaka05}
indicate a close spatial correlation between the stellar sources and
the surrounding material. Higher water ice and hydrocarbon absorptions in
the northern part of IRS$\,$13, as well as a redder continuum emission, suggest
that the IRS$\,$13N sources are more embedded in the ISM than the rest of the complex.
The radial velocity of $\eta$ is consistent with the radial velocity of the Bar
at this position (between 0 and 50$\,$kms$^{-1}$, \citealt{paum04}).
If the northern sources that show similar proper motions also have
correlated radial velocities, it would be reasonable to suggest that
the entire complex is indeed associated with the Bar. 
According to our orbital fits, IRS$\,$13N is located a few equivalent
arcseconds behind Sgr$\,$A*, which nicely agree with the view of \citet{paum04}
about the Bar being located somewhat behind Sgr$\,$A*.
This would imply that the 
IRS$\,$13E and IRS$\,$13N stellar complexes are spatially close. The existence of two comoving
clusters, each moving in a completely different direction, poses
a lot of challenges for star formation scenarios at the GC.

\subsection{Is IRS$\,$13N a bound system?}
\label{bound}
As can be seen from the orbital analysis, the best fit orbits for
IRS$\,$13N stars have rather different orbital parameters.
Observing six stars on different orbits
exactly at the moment when they appear to be very close in projection and 
to have similar proper motions is not very probable. 
There is no single orbit onto which all of the stars would
fit with a good $\chi$$^2$, but at least four of the stars
can be confined
to the same plane. In this case the stars are again on different
orbits and consequently have different
orbital periods, spanning the range from 1000 to 3000 years.  
This again implies that the  present
arrangement is temporary. Also, it is curious that the moment
at which we observe them 
coincides exactly with the presence of dense gas at their position.
Interestingly, their most likely common plane seems to coincide
with the plane of the counter-clockwise moving stars (CCWS; \citealt{paum06}).
The existence of the CCWS disk is still a matter of debate. According to \citet{paum06},
the disk is sparsely populated (12 stars), with $\eta$ belonging to it.
The assumption that all IRS$\,$13N stars belong to the CCWS significantly increases the
population of the disk, and at the same time weakens the claim that
the CCWS is essentially in non-circular motion. In the analysis of \citet{paum06},
seven out of twelve CCWS stars are on eccentric orbits ($e$$>$0.4), three of those 
belonging to IRS$\,$13E system.
We find that five of the IRS$\,$13N stars are on low-eccentricity orbits.

The analysis of 1$\sigma$ orbits for one star (Fig.$\,$\ref{orbits}$c$)
shows that the uncertainty of semi-major axes ($a$), as well as that of orbital periods ($P$),
is of the same order as the differences between $a$ and $P$ of each star. 
Therefore, at this point we still cannot exclude that IRS$\,$13N stars are indeed orbiting together. 
In the following we consider the possibility
that the cluster is bound.
We calculate the velocity dispersion of stars as 
\begin{displaymath}
  \sigma^2=\sigma^2_{pm} - \sum_{i=1}^N[error^2(v_{x,i}) + error^2(v_{y,i})]/[2(N-1)] 
\end{displaymath}
where $\sigma$$_{pm}$ is the dispersion of the measured proper motions and
the second term removes the influence of the proper motion measurement uncertainties. 
 
The mass estimate from the
velocity dispersion of stars gives $\sim$3300$\,$\solm,
an unrealistically high mass for such a cluster. 
For a comparison, we note that \citet{paum06} estimate
the total stellar mass of IRS$\,$13E to be $\sim$1000$\,$\solm.
We note that even if the uncertainties in proper
motion measurements were underestimated, the final mass
of the cluster calculated from the 
velocity dispersion would remain high. For instance, a hypothetical underestimation
of our uncertainties of 30$\%$ would result in $\sim$8$\%$ lower velocity dispersion and
the mass of 2700$\,$\solm.\\
The Hill radius 
\begin{displaymath}
r_{Hill} \approx a(1-e)\Big(\frac{m}{3M}\Big)^{\frac{1}{3}}
\end{displaymath}
for a cluster of $m$$\,$=$\,$3300$\,$\solm, with $a$$\,$=$\,$(2.8$\,$-$\,$5.7)'' 
and $e$$\,$=$\,$0.1$\,$-$\,$0.3
is $r_{Hill}$$\,$$\approx$$\,$0.17''$\,$-$\,$0.34'', with the 
average value remarkably close to 
the observed cluster radius of $\sim$$\,$0.25''.
The other way around, if we set $r_{Hill}$$\,$$\sim$$\,$0.25'', 
the maximum $a$ and minimum $e$,
the minimum mass
needed to keep the cluster from disruption would be 1250\solm.\\ 
The relaxation time for a system of $N$ stars with the mean mass $m$$_{*}$
is given by
\begin{displaymath}
t_{relax} \approx \frac{N}{8ln\Lambda}\times t_{cross}
\end{displaymath}
where $t$$_{cross}$=$R$/$v$ and $\Lambda$=$R$v$^2$/Gm$_{*}$$\approx$$N$
\citep{binneytremaine}.
Also, it is important to note that
tidal interactions tend to shorten relaxation timescales.
A cluster with the IRS$\,$13N velocity dispersion 
would have a $t$$_{relax}$ of the order of the IRS$\,$13N
orbital period only if it was consisted of 500-1000 stars.
For m$\sim$3300\solm contained in 500-1000 main sequence stars
we get that stars should be as bright as $m$$_K$=17-18 
(assuming the extinction $A$$_K$$\approx$3, \citealt{schoedel07}). These stars would
 be observed. For a lower mass limit of 1250\solm, we get $m$$_K$=19-21, exactly
at the limit of NACO K-band data.
Therefore it seems that 
the system could hardly be bound and survive for a significant time in 
the orbit around SgrA*.

If IRS$\,$13N stars are not bound, the observed velocity 
dispersion probably indicates that
the cluster is in the process of dissolution.
The velocity dispersion will subsequently diminish the
 stellar surface number density of the cluster, which would within
only a few hundred years
reach the background value given by \citet{schoedel07}.

\subsection{The nature of IRS$\,$13N}
The detailed discussion about the nature  of IRS$\,$13N sources is
presented in \citet{eckart04a}.
Among other possibilities, the authors argue that infrared 
excess sources IRS$\,$13N could represent
objects that have colors and luminosities consistent
with YSOs and Herbig Ae/Be stars. Dusty envelopes that usually surround
those objects would then give rise to the observed strong
infrared excess.
It is striking that six of the sources that are very close in
projection, also show very similar proper motion values.
In the previous section we argue that IRS$\,$13N could hardly survive
in the present arrangement for a significant amount of time.
The indication of a dynamically young stellar system 
 concurs with the \citet{eckart04a} hypothesis of IRS$\,$13N stars being
 extremely young objects. 
But why then does this object exsist at all? 
Timescales for its disruption seem to be much shorter
than any timescales within which stars normally form.
Here we must take into account that the Galactic Center 
is an extremely complex environment and that
any star formation process that takes place there is
completely different from the ``normal'' star formation
via the fragmentation of a molecular cloud. Timescales 
for star formation could be significantly shorter. The free-fall
time in a collapsing cloud at the distance of the young stars
in the central half parsec around Sg$\,$rA* is of the order 
of 100-1500$\,$yr. Also, simulations of star formation
in a gaseous disk around Sgr$\,$A* by \citet{nayakshin07}
indicate that the timescales for star formation 
are fairly short, of the order of several thousand years.
This is comparable to the orbital timescale of IRS$\,$13N.

A serious obstacle for forming stars at the GC
is that the gas densities in the central parsec
are way too low for the gas to be able to form stars. Even
the highest density estimates for the circumnuclear disk (CND; \citealt{christopher05})
are several orders of magnitude lower than required to fulfill
the criterion for Jeans instability. However, with 
the aid of shocks and collisions, a small clump of gas in-falling to the center could
be highly compressed and star formation within it triggered. 
As a matter of fact, the mini-spiral is a short-lived feature 
($t$$_{dyn}$$\sim$10$^{4}$$\,$yr) and is in-falling right now. Furthermore, the
material of the mini-spiral is susceptible to shocks, as discussed by e.g. \citet{muzic07}.
A possibility of forming stars in small associations has never been discussed so far
for this region. 
Star formation in small groups definitely would not be able to account for the
young co-eval stellar population in the central parsec, but could, in general, be possible
and result in small associations like IRS$\,$13N. 

However, it is clear that we still must be very careful 
when drawing a conclusion
that IRS$\,$13N cluster contains very young stars.
In case that the cluster is indeed a part
of the CCWS disk and that the stars were formed simultaneously with
other CCWS stars, then they must be several Myr old.
The fact that they are apparently  
strongly extincted may then indicate that
they are similar to
the mini-spiral bow-shock sources \citep{tanner05}, but
many times less massive and luminous (see discussion in \citealt{eckart04a}).
The question remains: why are they clustered?

\section{Conclusions}
\label{concl}

In this paper we present the proper motion
measurements of IRS$\,$13N, a newly discovered co-moving group of
red sources at the GC. We discuss the boundness of a cluster and 
a possibility that these stars are the youngest stellar object
ever observed in this environment.
Ongoing star formation in the GC would have significant implications 
on our understanding of this particular environment, and star formation
in galactic nuclei in general.
 It has been assumed that the
star formation in the GC occurred in two star bursts, $\sim$100$\,$Myr
and $\sim$7$\,$Myr ago \citep{krabbe95}. 
The existence of objects younger than 1$\,$Myr
 would imply that in-situ star formation in the GC and near massive
black holes in general is possible. Furthermore, this would also mean
that star formation in the GC must not necessarily be related to star bursts and
that, in fact, a continuous star formation could go on.


The question of the real
nature of the IRS$\,$13N cluster remains open.
More imaging and especially 
high resolution spectroscopic data that could
reveal features attributed to YSOs 
will be necessary to confirm our hypothesis
that the stars are extremely young.

\acknowledgements{We thank the referee, Dr. M. Morris, for comments
and suggestions that improved this work.
Part of this work was supported by the $Deutsche$ $Forschungsgemeinschaft$ (DFG) via SFB 494. 
K. Mu\v{z}i\'{c} and L. Meyer
 were supported for this research through a stipend from the International Max Planck Research School
(IMPRS) for Radio and Infrared Astronomy at the Universities of Bonn and Cologne.}
\bibliography{ms_IRS13N_4}

\begin{thebibliography}{41}
\expandafter\ifx\csname natexlab\endcsname\relax\def\natexlab#1{#1}\fi

\bibitem[{Baganoff {et~al.}(2003)Baganoff, Maeda, Morris, Bautz, Brandt, Cui,
  Doty, Feigelson, Garmire, Pravdo, Ricker, \& Townsley}]{bag03}
Baganoff, F.~K., Maeda, Y., Morris, M., {et~al.} 2003, ApJ, 591, 891

\bibitem[{Berukoff \& Hansen(2006)}]{berukoff06}
Berukoff, S.~J. \& Hansen, B. M.~S. 2006, ApJ, 650, 901

\bibitem[{Binney \& Tremaine(1987)}]{binneytremaine}
Binney, J. \& Tremaine, S. 1987, Galactic Dynamics (Princeton University Press)

\bibitem[{Brandner {et~al.}(2002)Brandner, Rousset, Lenzen, Hubin, Lacombe,
  Hofmann, \& Moorwood}]{brandner02}
Brandner, W., Rousset, G., Lenzen, R., {et~al.} 2002, Messenger, 107, 1

\bibitem[{Christopher {et~al.}(2005)Christopher, Scoville, Stolovy, \&
  Yun}]{christopher05}
Christopher, M.~H., Scoville, N.~Z., Stolovy, S.~R., \& Yun, M.~S. 2005, ApJ,
  622, 346

\bibitem[{Coker {et~al.}(2002)Coker, Pittard, \& Kastner}]{coker02}
Coker, R., Pittard, J.~M., \& Kastner, J.~H. 2002, A\&A, 383, 568

\bibitem[{Diolaiti {et~al.}(2000)Diolaiti, Bendinelli, Bonaccini, Close,
  Currie, \& Parmeggiani}]{diolaiti00}
Diolaiti, E., Bendinelli, O., Bonaccini, D., {et~al.} 2000, A\&AS, 147, 335

\bibitem[{Eckart \& Duhoux(1990)}]{eckart90}
Eckart, A. \& Duhoux, P. R.~M. 1990, in ASP Confrence Series, Volume 14,
  Astrophysics with Infrared Arrays (San Francisco: ASP), ed. R.~Elston, 336

\bibitem[{Eckart {et~al.}(2004)Eckart, Moultaka, Viehmann, Straubmeier, \&
  Mouawad}]{eckart04a}
Eckart, A., Moultaka, J., Viehmann, T., Straubmeier, C., \& Mouawad, N. 2004,
  ApJ, 602, 760

\bibitem[{Eisenhauer {et~al.}(2005)Eisenhauer, Genzel, Alexander, Abuter,
  Paumard, Ott, Gilbert, Gillessen, Horrobin, Trippe, Bonnet, Dumas, Hubin,
  Kaufer, Kissler-Patig, Monnet, Str{\"{o}}bele, Szeifert, Eckart,
  Sch{\"{o}}del, \& Zucker}]{eisenhauer05}
Eisenhauer, F., Genzel, R., Alexander, T., {et~al.} 2005, ApJ, 628, 246

\bibitem[{Eisenhauer {et~al.}(2003)Eisenhauer, Sch{\"{o}}del, Genzel, Ott,
  Tecza, Abuter, Eckart, \& Alexander}]{eisenhauer03}
Eisenhauer, F., Sch{\"{o}}del, R., Genzel, R., {et~al.} 2003, ApJ, 597, L121

\bibitem[{Genzel {et~al.}(2003)Genzel, Sch{\"{o}}del, Ott, Eisenhauer, Hofmann,
  Lehnert, Eckart, Alexander, Sternberg, Lenzen, Clénet, Lacombe, Rouan,
  Renzini, \& Tacconi-Garman}]{genzel03}
Genzel, R., Sch{\"{o}}del, R., Ott, T., {et~al.} 2003, ApJ, 594, 812

\bibitem[{Gerhard(2001)}]{gerhard01}
Gerhard, O. 2001, ApJ, 546, L39

\bibitem[{Ghez {et~al.}(2005)Ghez, Salim, Hornstein, Tanner, Lu, Morris,
  Becklin, \& Duchêne}]{ghez05}
Ghez, A.~M., Salim, S., Hornstein, S.~D., {et~al.} 2005, ApJ, 620, 744

\bibitem[{Hansen \& Milosavljevi\'{c}(2003)}]{hansen&milos03}
Hansen, B. M.~S. \& Milosavljevi\'{c}, M. 2003, ApJ, 593, L77

\bibitem[{Kim {et~al.}(2004)Kim, Figer, \& Morris}]{kim04}
Kim, S.~S., Figer, D.~F., \& Morris, M. 2004, ApJ, 607, L123

\bibitem[{Krabbe {et~al.}(1995)Krabbe, Genzel, Eckart, Najarro, Lutz, Cameron,
  Kroker, Tacconi-Garman, Thatte, Weitzel, Drapatz, Geballe, Sternberg, \&
  Kudritzki}]{krabbe95}
Krabbe, A., Genzel, R., Eckart, A., {et~al.} 1995, ApJ, 447, L95

\bibitem[{Lenzen {et~al.}(1998)Lenzen, Hofmann, Bizenberger, \&
  Tusche}]{lenzen98}
Lenzen, R., Hofmann, R., Bizenberger, P., \& Tusche, A. 1998, in Proc. SPIE
  Vol. 3354, Infrared Astronomical Instrumentation, ed. A.~M. Fowler, 606

\bibitem[{Levin(2007)}]{levin07}
Levin, Y. 2007, MNRAS, 374, 515

\bibitem[{Levin \& Beloborodov(2003)}]{l&b03}
Levin, Y. \& Beloborodov, A. 2003, ApJL, 590, L33

\bibitem[{Lu {et~al.}(2005)Lu, Ghez, Hornstein, Morris, \& Becklin}]{Lu05}
Lu, J.~R., Ghez, A.~M., Hornstein, S.~D., Morris, M., \& Becklin, E.~E. 2005,
  ApJ, 625, L51

\bibitem[{Maillard {et~al.}(2004)Maillard, Paumard, Stolovy, \&
  Rigaut}]{maillard04}
Maillard, J.~P., Paumard, T., Stolovy, S.~R., \& Rigaut, F. 2004, A\&A, 423,
  155

\bibitem[{McMillan \& Portegies~Zwart(2003)}]{mcm&p-z03}
McMillan, S. L.~W. \& Portegies~Zwart, S.~F. 2003, ApJ, 596, 314

\bibitem[{Milosavljevi\'{c} \& Loeb(2004)}]{milos&loeb04}
Milosavljevi\'{c}, M. \& Loeb, A. 2004, ApJ, 604, L45

\bibitem[{Moultaka {et~al.}(2005)Moultaka, Eckart, Sch{\"{o}}del, Viehmann, \&
  Najarro}]{moultaka05}
Moultaka, J., Eckart, A., Sch{\"{o}}del, R., Viehmann, T., \& Najarro, F. 2005,
  A\&A, 443, 163

\bibitem[{Mu{\v z}i{\'c} {et~al.}(2007)Mu{\v z}i{\'c}, Eckart, ~, Meyer, \&
  Zensus}]{muzic07}
Mu{\v z}i{\'c}, K., Eckart, A., ~, Sch{\"{o}}del, R., Meyer, L., \& Zensus, A.
  2007, A\&A, 469, 993

\bibitem[{Nayakshin \& Cuadra(2005)}]{nayakshin05}
Nayakshin, S. \& Cuadra, J. 2005, A\&A, 437, 437

\bibitem[{Nayakshin {et~al.}(2007)Nayakshin, Cuadra, \& Springel}]{nayakshin07}
Nayakshin, S., Cuadra, J., \& Springel, V. 2007, MNRAS, 379, 21

\bibitem[{Nayakshin {et~al.}(2006a)Nayakshin, Dehnen, Cuadra, \&
  Genzel}]{nayakshin06a}
Nayakshin, S., Dehnen, W., Cuadra, J., \& Genzel, R. 2006a, MNRAS, 366, 1410

\bibitem[{Nayakshin \& Sunyaev(2005)}]{nayakshin&sunyaev05}
Nayakshin, S. \& Sunyaev, R. 2005, MNRAS, 364, L23

\bibitem[{Paumard {et~al.}(2006)Paumard, Genzel, Martins, Nayakshin,
  Beloborodov, Levin, Trippe, Eisenhauer, Ott, Gillessen, Abuter, Cuadra,
  Alexander, \& Sternberg}]{paum06}
Paumard, T., Genzel, R., Martins, F., {et~al.} 2006, ApJ, 643, 1011

\bibitem[{Paumard {et~al.}(2004)Paumard, Maillard, \& Morris}]{paum04}
Paumard, T., Maillard, J.-P., \& Morris. 2004, A\&A, 426, 81

\bibitem[{Portegies~Zwart {et~al.}(2006)Portegies~Zwart, Baumgardt, McMillan,
  Makino, Hut, \& Ebisuzaki}]{p-z06}
Portegies~Zwart, S.~F., Baumgardt, H., McMillan, S. L.~W., {et~al.} 2006, ApJ,
  641, 319

\bibitem[{Reid \& Brunthaler(2004)}]{reid04}
Reid, M.~J. \& Brunthaler, A. 2004, ApJ, 616, 872

\bibitem[{Reid {et~al.}(2003)Reid, Menten, Genzel, Ott, Schödel, \&
  Eckart}]{reid03}
Reid, M.~J., Menten, K.~M., Genzel, R., {et~al.} 2003, ApJ, 587, 208

\bibitem[{Rousset {et~al.}(1998)Rousset, Lacombe, Puget, Hubin, Gendron, Conan,
  Kern, Madec, Rabaud, Mouillet, Lagrange, \& Rigaut}]{rousset98}
Rousset, G., Lacombe, F., Puget, P., {et~al.} 1998, in Proc. SPIE Vol. 3353,
  Adaptive Optical System Technologies, ed. D.~Bonaccini, 516

\bibitem[{Sch{\"{o}}del {et~al.}(2007)Sch{\"{o}}del, Eckart, Alexander,
  Merritt, Genzel, Sternberg, Meyer, Kul, Moultaka, Ott, \&
  Straubmeier}]{schoedel07}
Sch{\"{o}}del, R., Eckart, A., Alexander, T., {et~al.} 2007, astro-ph/0703178,
  accepted by A \& A

\bibitem[{Sch{\"{o}}del {et~al.}(2005)Sch{\"{o}}del, Eckart, Iserlohe, Genzel,
  \& Ott}]{schoedel05}
Sch{\"{o}}del, R., Eckart, A., Iserlohe, C., Genzel, R., \& Ott, T. 2005, ApJ,
  625, L111

\bibitem[{Stolte {et~al.}(2007)Stolte, Ghez, Morris, Lu, Brandner, \&
  Matthews}]{stolte07}
Stolte, A., Ghez, A.~M., Morris, M., {et~al.} 2007, arXiv:0706.4133v1

\bibitem[{Tanner {et~al.}(2005)Tanner, Ghez, Morris, \& Christou}]{tanner05}
Tanner, A., Ghez, A.~M., Morris, M.~R., \& Christou, J.~C. 2005, ApJ, 624, 742

\bibitem[{Zhao \& Goss(1998)}]{zhao&goss98}
Zhao, J.-H. \& Goss, W. 1998, ApJL, 499, L163

\end{thebibliography}
\end{document}